\documentclass[twocolumn,aps]{revtex4}
\usepackage{epsfig}
\usepackage[normalem]{ulem}
\usepackage[usenames]{color}

\newcommand{\beq}{\begin{eqnarray}}
\newcommand{\eeq}{\end{eqnarray}}

\usepackage[normalem]{ulem}
\usepackage[usenames]{color}

\begin{document}

\title{\boldmath{}Modeling $^{19}$B as a $^{17}$B-$n$-$n$ three-body system in the unitary limit}

\author{$^{1,2}$Emiko Hiyama}
\address{$^1$Department of Physics,  Kyushu University, Fukuoka,  819-0395, Japan}
\address{$^2$RIKEN Nishina Center, Wako 351-0198  Japan}

\author{$^{3}$Rimantas Lazauskas}
\address{$^3$IPHC, IN2P3-CNRS/Universit\'e Louis Pasteur BP 28, F-67037 Strasbourg Cedex 2, France}

\author{$^{4}$F.~Miguel Marqu\'es}
\address{$^{4}$LPC Caen, Normandie Universit\'e, ENSICAEN, Universit\'e de Caen, CNRS/IN2P3, F-14050, Caen, France}

\author{$^{5}$Jaume Carbonell}
\address{$^{5}$Institut de Physique Nucl\'eaire, Universit\'e Paris-Sud, IN2P3-CNRS, 91406 Orsay Cedex, France}

\date{\today}

\begin{abstract}
 We present a model description of the bound $^{19}$B isotope
 in terms of a $^{17}$B-$n$-$n$ three-body system where the two-body subsystems $^{17}$B-$n$
 and $n$-$n$ are unbound (virtual) states close to the unitary limit. The $^{19}$B ground state is
 well described in terms of two-body potentials only, and two low-lying resonances are predicted.
 Their eventual link with the Efimov physics is discussed.
 This model can be naturally used to describe the recently discovered resonant states in $^{20,21}$B.
\end{abstract}

\maketitle

\section{Introduction}

 The interaction of a low-energy neutron ($n$) with a nucleus of mass number $A$ (with spin-parity $J^{\pi}_A$)
 is a balance between the attractive folded neutron-nucleus potential $V_{nA}$ and the effective Pauli repulsion
 between the incident and target neutrons.
 For $A=1$, all the $n$-$A$ spin ($S=J_A\pm1/2$) and isospin ($T$) channels are attractive:
 this attraction manifests through the large and negative values of the singlet scattering lengths
 ($a^{nn}_0=-18.5$~fm and $a^{np}_0=-23.5$~fm) and by the existence of the $^2$H bound state in the only
 triplet $n$-$p$ ($^3S_1$) channel \cite{KRS_ADNDT_1991}.

 The first consequence of Pauli repulsion already manifests for $A=2$, in the neutron scattering on $^2$H in
 the $S=3/2^+$ channel. The scattering length in this channel is positive, $a_{3/2}=6.35$~fm, and reflects a
 strong repulsion, while the $S=1/2^+$ channel, despite having a positive scattering length $a_{1/2}=0.65$~fm,
 remains attractive accommodating the $^3$H bound state.
 For $A=3$ and $T=1$ both spin channels are repulsive with $a_0=5.2$~fm and $a_1=4.8$~fm.
 The same happens in the case of $n$-$^3$He (isospin mixture) with $a_0=5.9$~fm (its real part)
 and $a_1=3.1$~fm. 
 The only existing reaction with  $A=4$, $n$-$^4$He, is also repulsive with $a_{1/2}=2.61$~fm.

 When increasing the neutron number in the target, the valence neutrons start filling $p$
 (and higher angular momentum) orbitals, the Pauli principle becomes less constraining and the net balance
 is again attractive with the corresponding negative values of $a_S$,
 starting from the $^7$Li ground state $J^{\pi}=3/2^-$ with $a_2=-3.6$~fm.
 This is also true for $^8$He with $a_S\sim-3$~fm \cite{Johansson,AlFalou} and $^9$Li with $a_S\sim-14$~fm
 \cite{AlFalou}, although in these cases the $n$-$A$ total spin $S$ was not determined.

 For even higher numbers of neutrons, the virtual state of the $n$-$A$ system starts approaching the threshold.
 A spectacular consequence of this trend manifests with the $^{17}$B isotope, where the $n$-$^{17}$B virtual state 
 is located at the extreme vicinity of $^{18}$B ground-state threshold. 
 This effect is reflected by the largest value of the neutron-nucleus scattering length observed so far,
 $a_S\sim-100$~fm \cite{MSU_2012}. However, the limited resolution and acceptance of the experiment did not allow
 to fix a lower bound for the scattering length~\cite{MSU_2012}, and only an upper bound $a_S<-50$~fm was determined.

 Despite this experimental uncertainty, the potentially huge $n$-$^{17}$B scattering length confers a unique
 and intriguing character to $^{19}$B, a two-neutron Borromean halo nucleus~\cite{B17-19_RMS} exhibiting a weakly
 bound core-$n$-$n$ structure in which all the two-body subsystems are unbound~\cite{Borromean}.
 Moreover, due to its extremely weak binding ($2n$ separation energy of $S_{2n}=0.14\pm0.39$~MeV \cite{B19_mass}),
 $^{19}$B has no bound excited states.
 As such, the ground state can be seen as an extended three-body system resulting from the very large
 scattering lengths of its two two-body components (of the order of $-20$ and $-100$~fm respectively).

 The aim of this short note is to present a dynamical model describing simultaneously both structures,
 $n$-$^{17}$B scattering and the ground state of $^{19}$B, which could be easily applicable to the description
 of the heavier boron isotopes $^{20,21}$B recently observed in Ref.~\cite{20B_PRL121_2018}.
 The model is based on a two-body $n$-$^{17}$B local potential, built to this aim,
 supplemented with a realistic $n$-$n$ interaction.

 The same dynamical approach was used in Ref.~\cite{Mazumdar2000} to describe $^{19}$B. However, it overbound the
 ground state and predicted bound excited states due to the purely attractive $n$-$^{17}$B non-local interaction.
 Similar models, based on EFT renormalized zero-range $n$-core interaction plus a three-body force,
 were applied to another Borromean nucleus, $^{22}$C \cite{22C_Yamashita_2011,22C_Phillips_2013}.
 Although the large scattering length values involved in the $^{19}$B subsystems make natural its description
 in terms of EFT zero-range plus three-body forces, we have preferred in view of further applications to rely
 only on two-body finite-range interactions.
 A more consistent attempt to link the $n$-$A$ low-energy parameters (LEP) with the  reaction cross sections with the
 target nucleus was proposed in Refs.~\cite{Blanchon_ NPA739_2004,Blanchon_NPA784_2007}.

 Due to the experimental uncertainties in the relevant observables,
 respectively the $n$-$^{17}$B scattering length and $^{19}$B binding energy,
 we will explore the possible range of those parameters up to the unitary limit.
 Finally, our model predicts the first excited states of $^{19}$B to be of resonant character,
 in agreement with experimental observations~\cite{Gibelin_FB22}.

\section{Neutron-nucleus potential}

 It is assumed that a low-energy neutron approaching $^{17}$B feels a short-range repulsion due to the
 Pauli principle with respect to the 12 neutrons in the $^{17}$B core, plus a loose attraction due to
 the folded $n$-$A$ interaction. A simple form accounting for these facts can be:
\begin{equation}  \label{V}
  V_{n^{17}{\text{B}}}(r) \ = \ V_r \, \left( e^{-\mu r} - e^{-\mu R} \right) \, {e^{-\mu r} \over r}
\end{equation}
 where $R$ is a hard-core radius, hindering the penetration of the incoming neutron in the nucleus,
 and $\mu$ is a range parameter for the folded $n$-$^{17}$B potential.
 An educated value can be $R=3$~fm, which corresponds to the r.m.s.\ matter radius of $^{17}$B \cite{B17-19_RMS},
 and we take $\mu=0.7$~fm$^{-1}$ corresponding to the pion mass.
 
\vspace{10mm}
 \begin{figure}[h!] 
\begin{center}
\epsfig{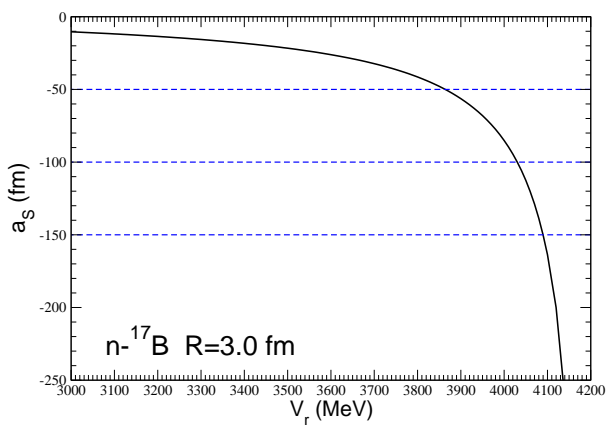}
\caption{Dependence of the $n$-$^{17}$B scattering length $a_S$ on $V_r$ for $R=3$~fm.
  The dashed lines correspond to some selected values of $a_S$ used in other figures.}\label{a0_Vr_299}
\end{center}
\end{figure}

 Once the range $\mu$ and the size $R$ are fixed, the potential depends on a single strength parameter $V_r$,
 which will be adjusted to reproduce the scattering length $a_S$.
 The numerical values along this work correspond to $m_n=939.5654$~MeV,
 $m_{^{17}{\text{B}}}=15879.1$~MeV, i.e.\ a $n$-$^{17}$B reduced mass $m_R=887.0771$~MeV
 and ${\hbar^2/2m_R}=21.9473$~MeV$\cdot$fm$^2$.

\begin{figure}[h]
\begin{center}
\epsfig{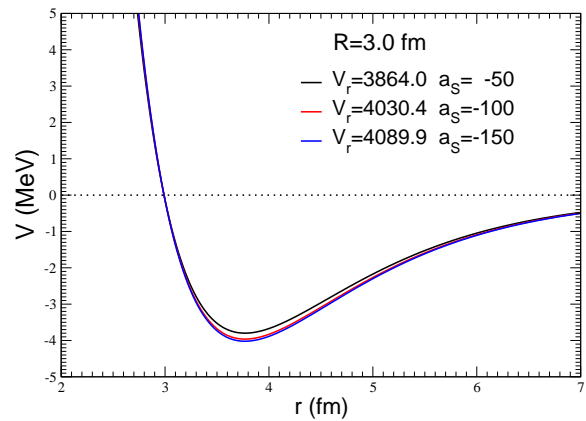}
\caption{The $V_{n^{17}{\text{B}}}$ potential for three different values of $V_r$ (in MeV) and the corresponding scattering lengths (in fm).}\label{V_a}
\end{center}
\end{figure}

 The dependence of the $n$-$^{17}$B scattering length $a_S$ on the strength parameter $V_r$ within this model
 is displayed in Fig.~\ref{a0_Vr_299}.
 The $n$-$^{17}$B potentials of Eq.~(\ref{V}) corresponding to $a_S=-50,-100,-150$~fm
 (dashed lines in Fig.~\ref{a0_Vr_299}) are displayed in Fig.~\ref{V_a}.
 As one can see, despite the large variation of $a_S$ the potentials quickly saturate when approaching
 the unitary limit. For example, when $a_S$ varies from $-50$ to $-100$~fm the potential changes by only
 a few tens of keV, and at this scale it looks almost independent beyond those values.
 
\vspace{.9cm}

 Until now we have ignored any spin-spin effect in the $n$-$^{17}$B interaction (spin symmetric approximation).
 In fact $^{17}$B is a $J^{\pi}=3/2^-$ state and can couple to a neutron in two different spin states, $S=1^-,2^-$.
 If the resonant scattering length is due to $S=2^-$ state, as it is assumed in \cite{MSU_2012}, there is
 no reason for the $S=1^-$ to be resonant as well. To account for this eventual asymmetry in further calculations,
 we have introduced  a spin-dependent interaction by assuming for each spin $S$-channel a potential $V_{n^{17}{\text{B}}}^{(S)}$ 
 having the same form  as Eq.~(\ref{V})  and driven by the corresponding strength parameter $V_r^{(S)}$.
 This can be cast in a single expression using  the standard  operator form:
\begin{equation}\label{Vss}
 V_{n^{17}{\text{B}}}(r) \ = \ V_c(r) + (\vec{S}_n\cdot\vec{S}_{^{17}\text{B}}) \, V_{ss}(r)
\end{equation}
in terms of the spin-spin operator:
\begin{equation}\label{SS}
 2(\vec{S}_n\cdot\vec{S}_{^{17}\text{B}})= S(S+1)- {9\over 2}  
 \end{equation}
and the central $V_c$ and spin-spin $V_{ss}$ components which are expressed in terms of $V_{n^{17}{\text{B}}}^{(S)}$  by inverting
the linear system obtained from Eq. (\ref{Vss}) for $S=1$ and $S=2$:
\begin{eqnarray}
4V_{n^{17}{\text{B}}}^{(S=1)} &=&4V_c - {5} V_{ss} \cr
4V_{n^{17}{\text{B}}}^{(S=2)} &=&4V_c + {3} V_{ss} 
\end{eqnarray}
It follows from that,  that  the central ($V_c$) and spin-spin ($V_{ss}$) components  have  the same form as Eq.~(\ref{V}):
\begin{eqnarray*}
 V_{i}(r) \ = \ V_r ^{(i)} \, \left( e^{-\mu r} - e^{-\mu R} \right) \, {e^{-\mu r} \over r} \qquad i=c,ss
\end{eqnarray*}
 with the strength coefficients given by:
\[ V_r ^{(c)} = \,{1\over8}\left(3V_r^{(1)}+5V_r^{(2)}\right) \hskip3mm 
   V_r^{(ss)} = \,{1\over2}\left( V_r^{(2)}- V_r^{(1)}\right) \]
 Using this $n$-$^{17}$B potential, the interest of our model lies in the description of the more complex
 systems composed by $^{17}$B and several neutrons in terms of two-body interactions.
 The first step in this direction is $^{19}$B, which will be the subject of the next section.

\section{$^{19}$B results}

 The $^{19}$B nucleus will be described as a $^{17}$B-$n$-$n$ three-body system. The $n$-$^{17}$B potential
 of Eq.~(\ref{V}), supplemented by a $n$-$n$ interaction, will constitute the three-body Hamiltonian.
 This description is physically justified by the large values of the scattering length in each two-body subsystem.
 
 For the $n$-$n$ potential, we have chosen two different models: the Bonn~A potential \cite{Bonn_A}
 and a charge dependent (CD) version of the semi-realistic MT13 interaction \cite{MT_NPA127_1969}.
 The Bonn~A potential, being charge independent, provides the $n$-$n$ low-energy parameters 
 $a_{nn}=-23.75$~fm and $r_{nn}=2.77$~fm and acts in all partial waves.

 We have built a CD version of MT13 starting from the original parameters of the NN singlet state 
 (that is $V_R=1438.720$~MeV$\cdot$fm, $\mu_R=3.11$~fm, $\mu_A=1.55$~fm):
\begin{equation}\label{V_nn}
 V_{nn} \ = \ V_R\, {e^{-\mu_R r} \over r} \, - \, V_A\, {e^{-\mu_A r} \over r}
\end{equation}
 and adjusting the strength of the attractive term to $V_A=509.40$~MeV$\cdot$fm in order to reproduce
 the $n$-$n$ LEP $a_{nn}=-18.59$~fm and $r_{nn}=2.93$~fm, in close agreement with the experimental values.
 In order to cross-check the results, the three-body problem was solved independently by using two different
 formalisms: Faddeev equations in configuration space \cite{LC_3n_PRC72_2005,FE_3B_20054} 
 and the Gaussian Expansion Method \cite{Hiya03}.

\begin{figure}[t]
\begin{center}
\epsfig{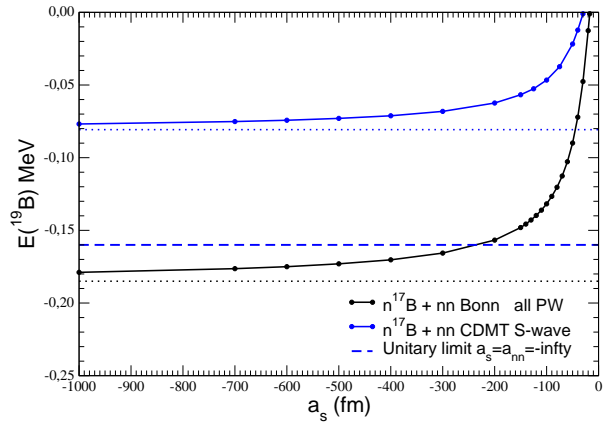}
\caption{$^{19}$B ground-state energy with respect to the first particle threshold as a function of $a_S$,
  for $R=3$~fm. Blue and black lines correspond, respectively, to model versions {\it(i)} and {\it(ii)}.}\label{E_19B_a}
\end{center}
\end{figure}

 We have first computed the $^{19}$B ground-state energy $E(^{19}$B) as a function of the scattering length
 $a_S$ in the spin-symmetric case, that is with $V_r^{(1)}=V_r^{(2)}=V_r$.
 Results, measured with respect to the $^{17}$B-$n$-$n$ threshold (i.e.\ $-S_{2n}$),
 are displayed in Fig.~\ref{E_19B_a}. They concern two different versions of the model:
 {\it(i)} a purely $S$-wave interaction both in $V_{n^{17}{\text{B}}}$ and in $V_{nn}$ (solid blue line) and
 {\it(ii)} letting the interaction of Eq.~(\ref{V}) act in all partial waves (solid black line)
           supplied with the Bonn~A model for $V_{nn}$.

 In view of these results, the following remarks are in order:
\begin{itemize}

 \item  The quantum numbers of $^{19}$B ground state are $L=0$ and $S=3/2$, which are separately conserved,
 and so $J^{\pi}(^{19}\text{B})=3/2^{-}$. {Notice that since the total angular momentum is $L=0$ the total parity is given by the intrinsic parity of $^{17}$B, which is a 
 $3/2^-$ state.}

 \item In the range of $a_S$ values compatible with the experiment ($a_S<-50$~fm \cite{MSU_2012})
 and in both versions of our model, $^{19}$B is bound.  
 Its binding energy decreases when $a_S$ increases and the binding disappears for a critical value of the
 scattering length $a_S^c$ which slightly depends on the model version: $a_S^c\approx-30$~fm for {\it(i)} and
 $a_S^c\approx-15$~fm for {\it(ii)}.

\item The $^{19}$B binding energy is compatible with the experimental value $E=-0.14\pm0.39$~MeV \cite{B19_mass}
 for both versions of the model and in all the range of $a_S$, starting from $a_S=-50$~fm until the unitary limit
 in the $n$-$^{17}$B channel, i.e.\ $a_S\to-\infty$.
 This limit (dotted lines) corresponds to $E_u=-0.081$~MeV in version {\it(i)} (blue)
 and $E_u=-0.185$~MeV in version {\it(ii)} (black). Note that in both cases the ground-state energy for
 $a_S=-150$~fm is only $\sim20$--30~keV distant from the unitary limit.

\item It is interesting to consider also the unitary limit in the $n$-$n$ channel, i.e.\ $a_{nn}\to-\infty$.
 This is realized, in the purely $S$-wave model version {\it(i)}, by setting $V_A\approx531.0$~MeV in the
 $n$-$n$ potential of Eq.~(\ref{V_nn}).
 The full unitary result, where both $a_S=a_{nn}\to-\infty$, is indicated by the dashed blue line in this figure.
 It corresponds to a $^{19}$B energy $E_{uu}=-0.160$~MeV, compatible with the experimental result.
 The {\it ab-initio} nuclear physics in the $S$-wave unitary limit of the NN interaction was recently considered
 in Ref.~\cite{Unitary_NP_PRL118_2017}. It was claimed that the gross properties of the nuclear chart
 were already determined by very simple NN interactions, provided they ensure $a_0=a_1\to-\infty$.
 The $^{19}$B ground state constitutes a nice illustration of this remarkable property,
 though at the level of cluster description and based on other dynamical contents.

\item The main difference between version {\it(i)} and {\it(ii)},
 solid blue and black lines in Fig.~\ref{E_19B_a} respectively,
 is essentially due to the contribution of the higher angular-momentum terms in $V_{n^{17}{\text{B}}}$.
 In the present state of experimental knowledge, this contribution is totally uncontrolled but we have made
 an attempt to quantify it by assuming $V$ to be the same in all partial waves.
 For $a_S=-150$~fm this accounts for about 60~keV of extra binding.
 The difference due to the $n$-$n$ interaction is smaller, and mainly given by the fact that Bonn~A model is
 charge independent and has a larger $a_{nn}$ absolute value.
 Once this is corrected, the difference in the $^{19}$B binding energy at $a_S=-150$~fm reduces to 8~keV,
 which is mainly accounted for the higher partial waves in the $n$-$n$ interaction.
 As expected from the Effective Field Theory arguments, the details of the interaction turn out to be
 negligible for a system close to the unitary limit.

\end{itemize}

\begin{figure}[t]
\begin{center}
\epsfig{file=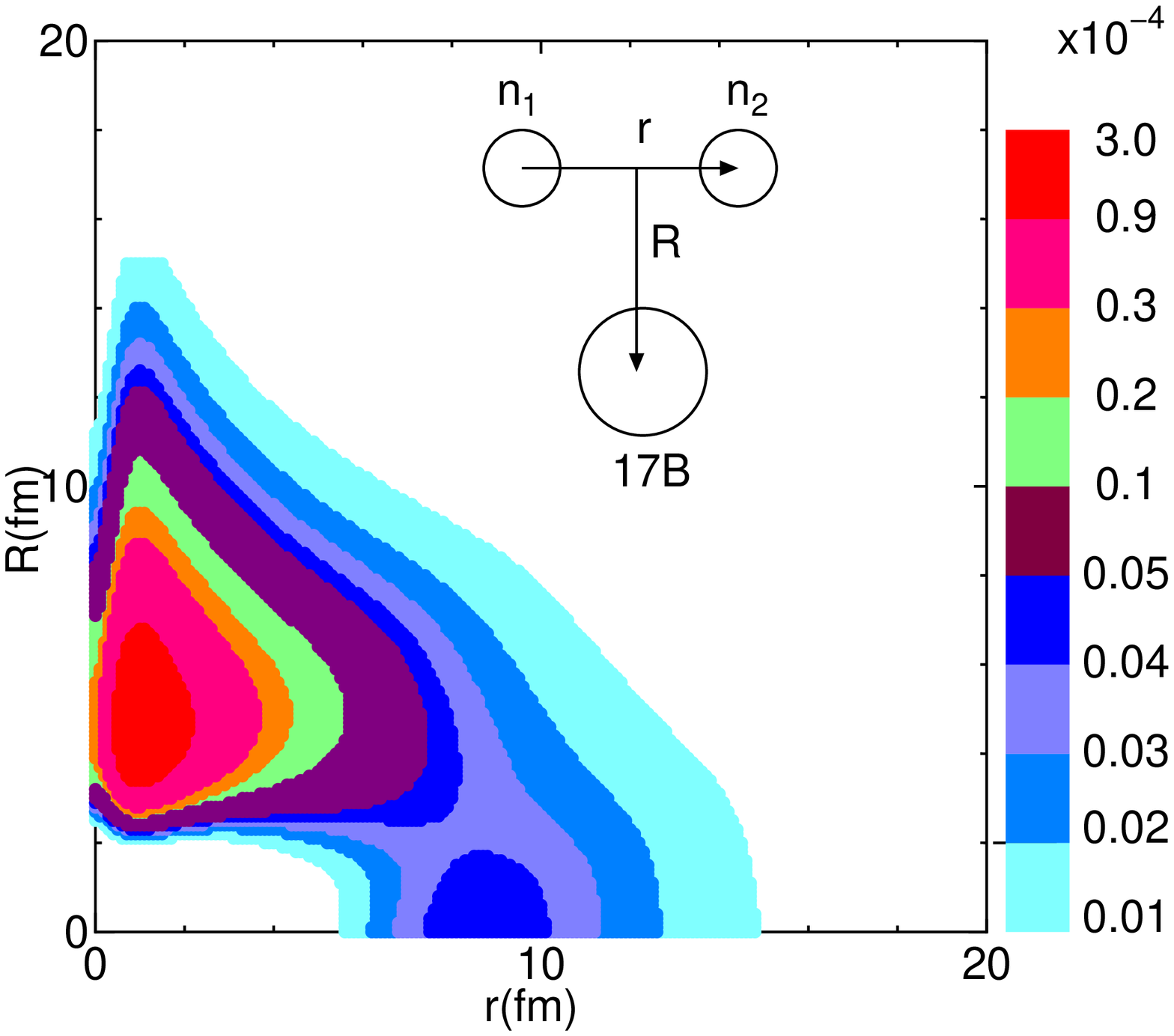,scale=0.33}
\epsfig{file=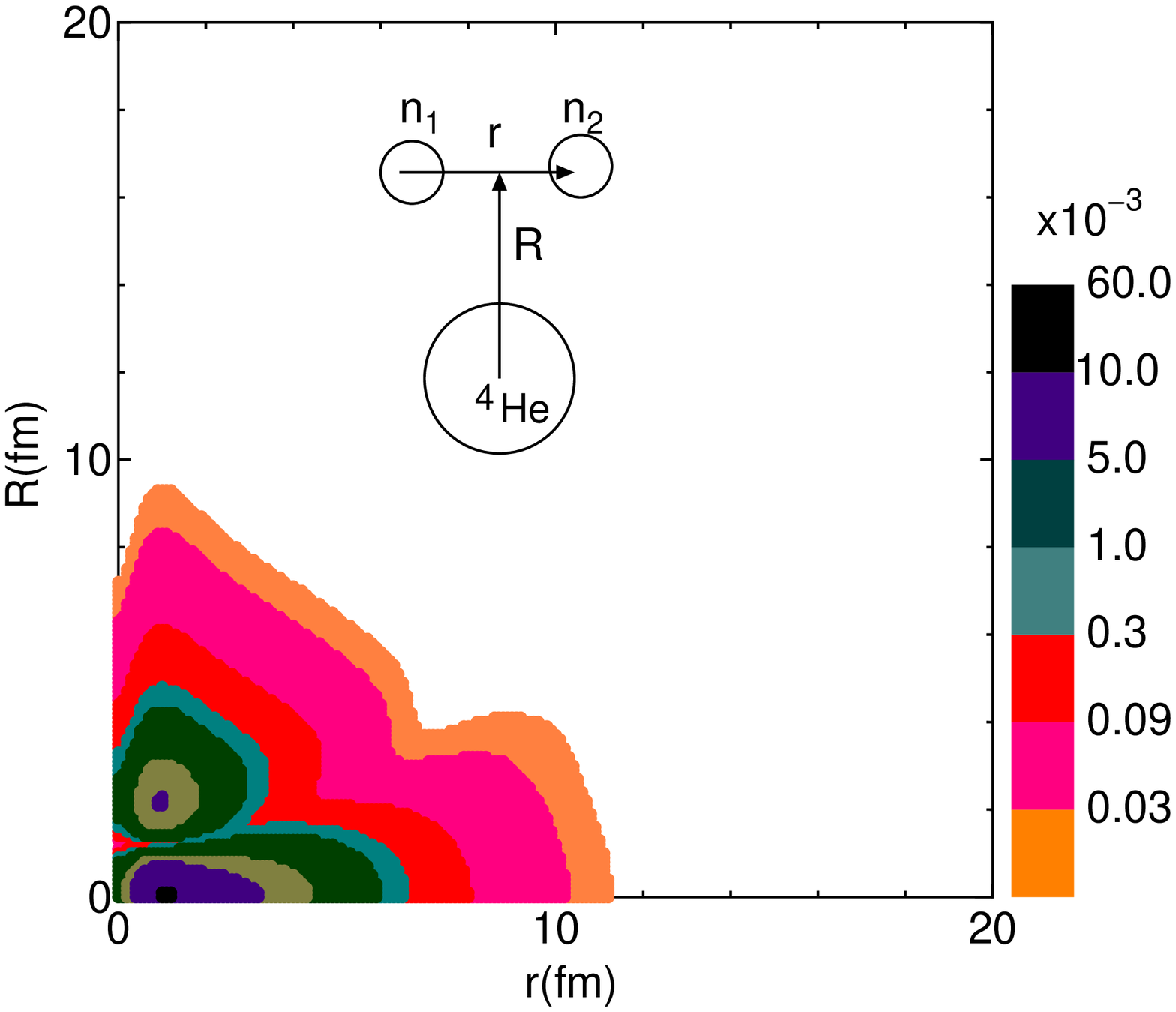,scale=0.33}
\caption{Ground-state probability amplitude $|\Psi(r,R)|^2$ as a function of the Jacobi coordinates:
  upper panel $^{19}$B for $a_S=-100$~fm in model version {\it(ii)} and
 lower panel for $^6$He~\cite{Hiyama_he6}.}\label{Density}
\end{center}
\end{figure}

The spatial probability amplitude, i.e.\ the squared modulus of the total wave function $\Psi(r,R)$ in terms
 of the Jacobi coordinates, is represented in Fig.~\ref{Density} (upper panel),
 for $a_S=-100$~fm and $E=-0.130$~MeV in model version {\it(ii)}.
 For the sake of comparison, we have computed the same amplitude and in the same scale of another two-neutron halo nucleus, $^6$He ($E=-0.97$~MeV),
 using the $\alpha$-$n$-$n$ three-body model from Ref. ~\cite{Hiyama_he6}. It is displayed in the lower panel of the same Fig.~\ref{Density}.
 Due to the very weak binding energy, the wave function of  $^{19}$B is much more extended.
 Unlike $^6$He wave function it displays very large asymmetry, being strongly elongated in the $^{17}$B-$(nn)$ direction.
An exceptional feature of $^{19}$B is the presence of the void in the few fm space around the center of mass  of the three clusters $^{17}$B-$n$-$n$. 
This feature establishes  $^{19}$B  as a veritable two-neutron halo nucleus having a molecule-like structure with three centers. 

\bigskip
 Besides providing a satisfactory description of the $^{19}$B ground state, our model accommodates two broad
 resonances. Letting the interaction of Eq.~(\ref{V}) act in all partial waves with $a_S=-150$~fm
 ($V_r$=4089.9~MeV$\cdot$fm) and adopting the $S$-wave model for the $n$-$n$ interaction, two resonant states
 are found for total angular momentum $L=1$ and $L=2$, with energies
 $E_{1^-}\approx0.24(2)-0.31(4)i$~MeV and $E_{2^+}\approx 1.02(5)-1.22(6)i$~MeV.
 {The parameters of these resonances  however depend on the value of $a_S$,  which remains  unknown.
Their width display also a strong dependence on the size parameter ($R$) of the $n$-$^{17}$B interaction, that we fixed once at all to $R=3$ fm:
 if $R$ is increased, keeping $a_S$ fixed, the resonance widths decrease. Further experimental data are needed for a fine tuning of these parameters.}

 In fact, several resonances in the continuum of $^{19}$B have been recently observed \cite{Gibelin_FB22},
 although their precise energy and quantum numbers have not been determined yet.
 It is worth noting that despite its simplicity, our model of the $n$-$^{17}$B interaction is able to account for
 the $^{18}$B virtual state, the $^{19}$B ground state and two resonances without any need of three-body forces.
{This follows from the strong resonant character of  both $n$-$^{17}$B and $n$-$n$ channels
as well as the large  spatial extension of the  $^{19}$B ground state}.

\begin{figure}[t]
\begin{center}
\epsfig{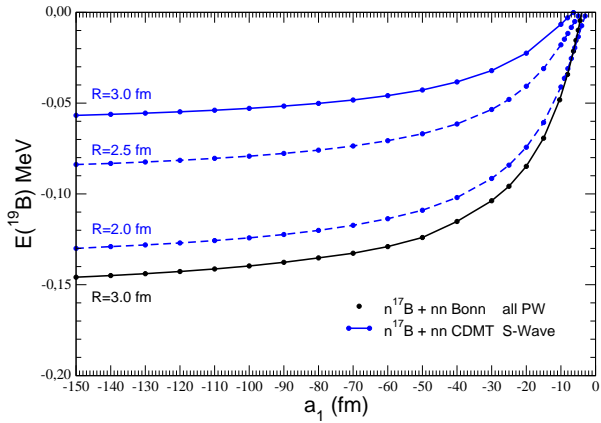}
\caption{$^{19}$B ground-state energy with respect to the first particle threshold as a function of $a_1$,
  for a fixed $a_2=-150$~fm and several values of $R$.
  Blue and black lines correspond, respectively, to model versions {\it(i)} and {\it(ii)}.}\label{E_19B_a_SSB}
\end{center}
\end{figure}

\bigskip
 If we introduce the spin dependence of the interaction, Eq.~(\ref{Vss}),
 and assume the non-resonant scattering length $a_1$ to be smaller in absolute value than $a_2$,
 the weaker potential leads to $^{19}$B binding energies smaller than those in Fig.~\ref{E_19B_a}.
 For example, if we fix $a_2=-150$~fm the $^{19}$B binding energy decreases when $a_1$ varies from the
 spin-symmetric case $a_1=a_2=a_S$ up to some critical value $a_1^c$, still negative, beyond which $^{19}$B
 is no longer bound. The results are displayed in Fig.~\ref{E_19B_a_SSB}.

 The very existence of this critical value $a_1^c$, as well as its negative sign, is independent of the
 model version (blue and black solid lines) and the size parameter $R$, that was also varied here
 to check the robustness of this result (dashed blue lines).
 In all the cases that we have explored, its value lies within $-7<a_1^c<-3$~fm (Fig.~\ref{E_19B_a_SSB}).
 On the other hand, in order to reach the critical value $a_1^c$ an extremely strong spin-spin
 interaction is required with $V^{(1)}_r/V^{(2)}_r\sim0.6$, for the case $a_2=-150$~fm.
 We may hardly find any physical arguments to support the existence of such a strong spin-dependence in the
 $n$-$^{17}$B interaction. 
 Therefore, we conclude that the value of $a_1$ should be also negative and not small,
 and that $^{19}$B remains bound even when the spin symmetry is broken.

\bigskip
 The proximity of the $n$-$^{17}$B interaction to the unitary limit strongly suggests that $^{17}$B could be
 a genuine nuclear candidate to exhibit the Efimov  physics \cite{NaidonEndo}, that is the existence of a family of bound states whose consecutive
 energies are scaled by a universal factor $f^2$. This is what would happen by setting  $a_1=a_2=a_{nn}=-\infty$  as in the blue dashed line of Fig.
 \ref{E_19B_a}. However the $^{17}$B-$n$-$n$ system, representing a light-light-heavy
 structure, turns out to be quite an unfavorable
 case to exhibit the sequence of excited Efimov states due to the requirement of a very large factor $f$.
 However, in the real world, as well as in our model, $^{19}$B has
 only one bound  state and it is governed by three different
 scattering lengths, from which only the $n$-$n$ scattering length is relatively well known and can be fixed to its experimental
 value. For the case when only the $n$-$^{17}$B interaction
 is tuned the universal factor turns out to be $f\approx2000$~\cite{NaidonEndo,Hove_Efimov}.
 It follows that the appearance of the first $L=0^+$ excited state of Efimov nature in
 $^{19}$B would manifest only when the $n$-$^{17}$B scattering length reaches several thousands of fm.
 We conclude that, independently of the particular value of $a_S$, it is highly unlikely to observe any Efimov
 excited state in $^{19}$B.
   On the other hand, the universal features related to Efimov physics~\cite{Mario} are however
 genuinely preserved in this system.
 
 It is worth mentioning that an hypothetical molecular-like $^{17}$B-$^{17}$B-$n$ system would constitute a
 favorable heavy-heavy-light structure in order to generate Efimov states~\cite{NaidonEndo}. However, the
 long-range part of the effective hyper-radial interaction in this system would be dominated by the repulsive
 Coulomb interaction between the $^{17}$B nuclei, largely exceeding the $1/r^2$ tail of Efimovian attraction.

\section{Summary}

 Motivated by the recent experimental results on heavy boron isotopes \cite{MSU_2012,Gibelin_FB22,20B_PRL121_2018},
 we have developed a model to describe the main phenomenological facts of these neutron-rich nuclei.
 It is based on a $n$-$^{17}$B effective interaction which, supplemented with a realistic $n$-$n$ potential,
 provides a satisfactory description of $^{19}$B ground and resonant states in terms of a $^{17}$B-$n$-$n$
 three-body system.

 The key ingredient of the model is the proper description of the extremely shallow virtual state of $^{18}$B,
 which manifests through a very large and negative value of the $n$-$^{17}$B scattering length, $a_S<-50$~fm,
 in the $S=2$ channel \cite{MSU_2012}.

 It is worth noticing that despite its simplicity, the model is able to account for the $n$-$^{17}$B virtual state,
 the $^{19}$B ground state and two broad resonances without introducing any three-body force.
 {Such a possibility is certainly due to the strong resonant character of the $n$-$^{17}$B and $n$-$n$ channels
 as well as the small binding energy and large spatial extension of  $^{19}$B}.
 In the spin-independent approximation, $^{19}$B is bound for $a_S<-30$~fm. Its binding energy is compatible
 with the experimental value for all the range of $a_S$ until the unitary limit $a_S\to -\infty$.

 We have considered the stability of our $^{19}$B results when a spin-spin interaction term is introduced in
 $V_{n^{17}{\text{B}}}$. By fixing a resonant value in the $S=2$ channel ($a_2=-150$~fm), we found that the
 system remains bound for a large domain of the $S=1$ scattering length $a_1<a_1^c<0$.
 The existence of this negative critical value $a_1^c$, beyond which the binding disappears,
 is independent of the model parameters, and its particular value depends only slightly on them.
 One has typically $a_1^c\sim-5$~fm. Such a large asymmetry of $a_S$, however, requires unphysically
 large differences among the potential strengths of the two spin channels.

Work is in progress to extend the application of this model to describe the recently observed resonances
 in $^{20}$B and $^{21}$B \cite{20B_PRL121_2018} in terms of $^{17}$B+$3n$ and +$4n$, respectively, by solving
 the corresponding four- and five-body equations using the techniques developed in Ref.~\cite{LHC_PLB791_2019}.
 On the other hand, the precise measurement of the main experimental observables related to this problem,
 such as the $^{18}$B scattering length, the $^{19}$B binding energy,
 and the energy and quantum numbers of the first $^{19}$B resonant states,
 would provide a more stringent constraint to the parameters of the present model,
 {and could eventually  suggest the introduction of $n$-$^{17}$B $P$-waves and/or $^{17}$B-$n$-$n$ three-body forces}.

\bigskip

\section*{Acknowledgments}

{The authors thank Val\'erie Lapoux and Anna Corsi for helpful discussions on the experimental $n$-$A$ scattering lengths.}
We were granted access to the HPC resources of TGCC/IDRIS under the allocation 2018-A0030506006 made by GENCI (Grand Equipement National de Calcul Intensif).
This work was supported by French IN2P3 for a theory project ``Neutron-rich light unstable nuclei''
and by the Japanese  Grant-in-Aid for Scientific Research on Innovative Areas (No.18H05407).

\end{document}